  \documentclass[aps,pre,groupedaddress,nofootinbib]{revtex4}
\usepackage{ifpdf}
\ifpdf
\fi

\usepackage{graphicx}
\usepackage{float}

\usepackage{hyperref}            

\begin{document}

%\preprint{}

\title{Methods for analytically estimating the resolution and intensity of neutron time-of-flight spectrometers.\\
The case of the TOFTOF spectrometer}


\author{Ana M Gaspar}
\email[]{ana.gaspar@frm2.tum.de}
\homepage[]{www.ph.tum.de/~agapsar}
%\thanks{}
%\affiliation{}


%\date{\today}
%\date{27th June 2005}
\begin{abstract}
  An analytical method is presented with allows to estimate the energy resolution of time-of-flight neutron spectrometers, as well
  as its partial contributions, over a dynamical range that extends from the elastic line to the accessible inelastic regions. 
  Such a method, already successfully applied in the past to the TOSCA and HET neutron inelastic scattering spectrometers 
  installed at the ISIS neutron spallation source [A M Gaspar, PhD Thesis, Universidade Tecnica de Lisboa, 2004], 
  is here applied to the high resolution
  time-of-flight spectrometer TOFTOF, mainly dedicated to quasi-elastic neutron scattering studies and
   installed at the new neutron reactor FRM II.  
  To make such calculations easily understandable, 
  the principle of work of the TOFTOF instrument and of each of its components is explained in detail.
  A simply method that can be used to estimate the instrument intensity, i.e. of the number of neutrons arriving at the sample 
  position per unit time, is also briefly outlined.
 
  To the benefit of the TOFTOF users, graphs displaying the dependencies of the instrument resolution at the elastic line and of 
  the instrument intensity on the relevant 
  instrument parameters, i.e the wavelength of the incident neutrons, the choppers speed of rotation and the frame overlap ratio, 
  are presented, in the form of iso-resolution or iso-intensity lines. 
  The method of estimation of the frame overlap ratio that is commonly used at time-of-flight instruments 
  such as TOFTOF is also explained and alternative options concerning this parameter, depending on the dynamical range of interest, 
  are briefly addressed. 
\end{abstract}

\pacs{}
%\keywords{}

\maketitle

\vskip 50pt
\tableofcontents
 

%\section{ola }


%\subsection{}
%\subsubsection{}
\newpage
\section{Description of the TOFTOF spectrometer}
	The TOFTOF spectrometer \cite{Zirkel_2000,Zirkel_report,Roth_thesis,toftof_webpage} is constituted by two 
	\underline{t}ime-\underline{o}f-\underline{f}light sections,
	where advantage is taken from the fact 
	that (unlike electromagnetic waves) 
	neutrons with different energies travel at different velocities, and hence take 
	different times to travel a fixed distance - {\it L}.	
	The primary 
	section consists of a set of disc choppers, whose rotation around an horizontal axis parallel 
	to the direction of the neutron beam 
	allows to obtain a pulsed monochromatic beam at the sample position, the 
	energy of the neutrons
	incident on the sample being basically determined by the phase difference between the first and 
	the last pairs of choppers, located L$_0$=10 m apart. %(figure \ref{figure1}). 
	The secondary part of the spectrometer consists of an argon filled flight chamber
	forcing the scattered neutrons to travel a distance L$_1$=4 m between the sample and 
	the detectors and allowing for the determination of their energy by measuring the time they
	take to travel that distance.
	
	\begin{figure}[H]
      	\centerline
	  {
	    \ifpdf
	    \includegraphics[scale=2.0]{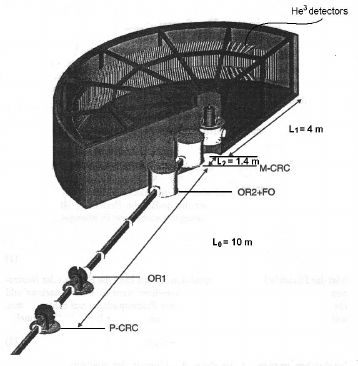}
            \else
	    \includegraphics[scale=2.0]{TOFTOF_scheme.eps}
	    \fi
          }			
	  \caption{\small \label{figure1} Schematic representation of TOFTOF spectrometer. Figure from
	\cite{Zirkel_report} (adapted).}
        \end{figure}

  \subsection{The neutrons seen by TOFTOF}
	The instrument is located in the neutron-guide hall of the FRM II reactor \cite{frm2_webpage} receiving 
	moderated neutrons from its liquid deuterium cold source (T=25K), through an S-shaped neutron 
	guide (of cross-section 44$\times$100 mm$^2$ and radius $\rho$=2000 m). The use of a curved
	S-shaped guide, which cuts-off neutrons of wavelengths smaller than 
	1.4\AA, ensures that the chopper system will not be 
	irradiated with neutrons too energetic to be fully absorbed by the necessarily thin 
	gadolinium coating of the choppers \cite{Roth_2000}. 
        Because of that and of the aluminium windows the neutron beam must cross,
        %Therefore, at the sample position, 
        the Maxwellian distribution defining the 
        neutron flux per unit wavelength appears slightly distorted \cite{TOFTOF_report2004}.
	% hence allowing to choose 
	%the incident pulse wavelength over the range 2\AA - 12\AA$\:$ 
	%(or energies in the range 0.5 - 20 meV).
	%Figure 2 represents the neutron flux at the sample position of the TOFTOF spectrometer.
 
	%In figure 2 the measured 
	%incident neutrons spectrum at the TOFTOF sample position is displayed.
	 
	%\begin{figure}[H]
      	%\centerline
	%  {
	%    \ifpdf
	%    \includegraphics[scale=2]{Neutron_flux.pdf}
        %    \else
	%    \includegraphics[scale=2]{Neutron_flux.eps}
	%    \fi
        %  }			
	%  \caption{\small \label{figure2} Differential neutron flux at the TOFTOF sample position
	%(beam cross section: 23 mm $\times$ 46 mm, integrated flux: $1\times10^{10}$ n/cm$^2$/s).
	%Figure reproduced from \cite{TOFTOF_report2004}. 
	%}
    %\end{figure}	
  
  \subsection{The chopper system}

	The chopper system is composed in total by seven 
	carbon fiber composite chopper discs of diameter 600 mm,
	which can rotate at speeds ranging from 3000 up to 27000 rpm (50 - 450Hz).
	Each of these chopper discs contains one or several slits, through which
	neutrons will be able to pass, whereas neutrons hitting the other parts of the 
	chopper discs will be absorbed.
 
	The first two chopper discs form a counter-rotating (CR) pair, 
	responsible for pulsing the incident beam, 
	while the last two (also forming a CR
	pair of chopper discs) will monochromatize the neutron pulses, by choosing a specific 
	time delay of their opening time with regards to the opening time of the pulsing pair of 
	choppers.
	%For instance, in order to select the 5 meV neutrons from
     	%a white pulse leaving the pulsing choppers at the instant $t_0$, the monochromating
	%choppers placed 10 m away would need to be at the maximum transmission position at 
        %t=$t_0$+$1\times10^{-2}\;$s (numerical value obtained from expression \ref{tflight}).
	The main advantage of the use of a CR pair of choppers is that of allowing 
	for a doubling of the transmitted intensity for the same opening time, when compared
	to single choppers (since, given the doubling of the relative speed of the slits, the 
	CR choopers slit width may be the double of that of a single chopper)\cite{Copley1988,Copley1991}. 
  
       \begin{figure}[H]
      	\centerline
	  {
	    \ifpdf
	   \includegraphics[scale=1.0]{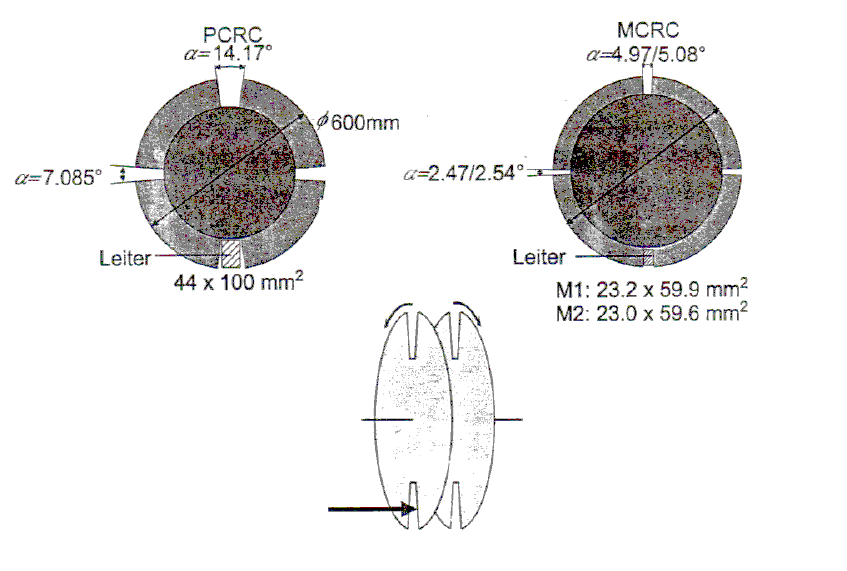}
            \else
	    \includegraphics[scale=1.0]{TOFTOF_choppers.eps}
	    \fi
          }			
	  \caption{\small \label{figure4} Schematic representation of TOFTOF's Pulsing and
	Monochromating choppers with indication of the widths of their slits. Figure from \cite{Zirkel_report}. 
	}
    	\end{figure}	
	 The time FWHH of the 
	%Gaussian equivalent to the 
	transmission function of the counter-rotating pair of choppers
	may be approximately expressed by\protect\footnote{
          %\footnotesize{
	The time uncertainties, are in practice commonly 
     defined by the full width at half-height (FWHH) of the time distributions. 
      A generally used method, which was 
     also applied in the analysis here presented, 
	is to approximate the distributions to Gaussians with the 
     same root mean square deviation. In this case, the FWHH can be obtained analytically and is simply
     given by $\Delta t = \sqrt{8{\rm ln}(2)}\;\sigma_{t}$. Gaussians have the useful property that 
     they may be convoluted simply by taking the root of the sums of the squares of their widths (
     $\sigma _t=\sqrt{\sigma^2_{t1}+\sigma^2_{t2}+...}$),
     %\begin{equation}
     %  \sigma _t=\sqrt{\sigma^2_{t1}+\sigma^2_{t2}+...}
     %  \label{convolution}
     %\end{equation}
     which avoids the need to perform the convolution integrals and gives very similar final results.\\ 
     In the case of a triangular distribution as the one expected to be the transmission function
	of the rotating choppers, one has 
	$\sigma_{\triangle}={FWHM}_{\triangle}/\sqrt{6}$ and hence the FWHM of the equivalent Gaussian 
	coincides with the FWHM of the transmission function since 
	%${FWHM}_{Gauss}\simeq {FWHM}_{\triangle}$ since 
	$\sqrt{8\ln{2}/6}\simeq 1$.
	Imperfections in the collimators that define the slit width, the non-infinitesimal distance 
	between the two counter-rotating choppers and the finite transmission of 
	neutrons through the absorbing layers tend to make the actual transmission functions assume a 
	more Gaussian shape. In some situations, the root mean square deviation 
	of the real transmission function was found to be better expressed by 
     	$\sigma=FWHH/2\sqrt{\ln 2}$ \cite{windsor}, in which case, the gaussian equivalent would be
	${FWHM}_{Gauss}\simeq \sqrt{2}{FWHM}_{trans}$.
	Previous ray-tracing simulations also seem to indicate this to be the case at TOFTOF instrument
	\cite{Roth_thesis}, which was then taken into account in the resolution 
	calculations here presented.
	%In the resolution calculations here presented the FWHM of the Gaussian equivalent to the 
	%transmission function was for the moment 
	%assumed to be equal to the one given by expression \ref{chopper_fwhm}.
	%}
        } 

	\begin{equation}
		\Delta t =\frac{b}{2 \pi f D} 
		\label{chopper_fwhm}
	\end{equation}
	where $b$ represents the choppers slit width, $f$ the choppers frequency of rotation and 
	$D$ the choppers diameter.
	%hence requiring high chopper frequencies for sharp transmission functions.
	
	%\begin{figure}[H]
      	%\centerline
	%  {
	%    \ifpdf
	%    \includegraphics[scale=0.25]{squemes2.pdf}
        %    \else
	%    \includegraphics[scale=0.25]{squemes2.eps}
	%    \fi
        %  }			
	%  \caption{\small \label{figure3} Schematic representation of the TOFTOF's chopper system. 
	%}
    	%\end{figure}

	Both the pulsing and the monochromating pairs of choppers contain four slits located
	at 90$^\circ$ as displayed in figure 4. 
	Of the neutron
	pulses produced by passing through each of these two pairs of opposite 
	slits only the ones created by one of them (either
	the larger or the smaller ones) are allowed to proceed to the sample, by means of the 
	use of two other chopper discs located after the pulsing pair of choppers and before the 
	monochromating pair of choppers. These two choppers also guarantee 
	the uniqueness of the energy to be selected by the monochromatoring
	pair of choppers (and for that reason are called Order Removal choppers).

	%Of the other three chopper discs two of them guarantee that only one type of 
	%neutron pulses (short or large) proceeds to the sample  
	Finally, a seventh chopper, located  in between the order removal choppers 
	defines the so-called frameoverlap ratio, i.e.,
	the portion of pulses
	generated by the first pair of choppers that are 
	allowed to proceed to the last pair of choppers, and 
	hence to the sample (1, 1/2, 1/3, 1/4, ....),
	by spinning at a speed that is a fraction of the speed of the other 
	choppers.
	This last chopper is then the chopper that defines the repetition rate of the neutron 
	pulses impinging on the 
	sample and hence the maximum time window for analyzing the scattering of neutrons from one pulse,
	as a function of time-of-flight (hence of energy transfer), before the scattering from 
	another pulse comes into play.

	%From what was said above we can see that
	For instance, choppers spinning at 15000 rpm generate
	500 neutron pulses per second, the time separating two neutron pulses being therefore 2 ms.
	Assuming that this time window of observation is centered around the elastic line,
	if using incident neutrons of 5 meV ($\lambda$=4\AA), one would only be able to obtain the 
	spectrum of 
	the neutrons scattered with energies between 3.2meV and 8.7 meV (hence corresponding
	to the range from -3.7023 to +1.7 meV in energy transfer). If, however, one would 
	increase the time of observation to 6 ms by choosing a frameoverlap ratio equal to 3 
	(and hence disregarding every two out of three neutrons), one would already 
	be able to count 
	the neutrons scattered with energies between 1.65 meV and 69 meV 
	(hence from -64 meV to +3.35 meV in energy transfer), at the expense of a loss in the 
	elastic line intensity by a factor three.
	%For further illustration, the spectral coverages of acquisitions performed 
	%with different 
	%frame overlap ratios for different wavelength and chopper speed configurations are also
	%presented in table 1.
	        
\begin{figure}[H]
      	\centerline
	  {
	    \ifpdf
	    \includegraphics[scale=0.28]{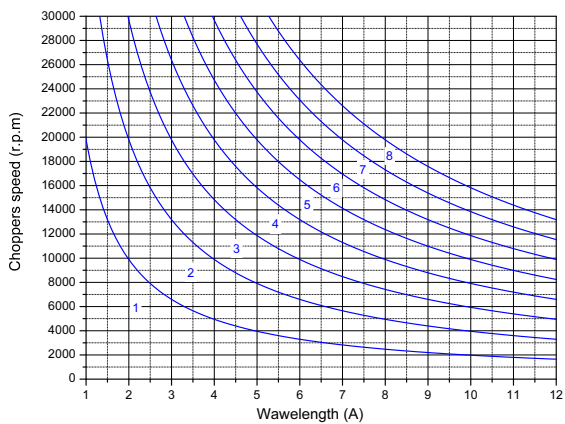}
            \else
	    \includegraphics[scale=0.30]{TOFTOF_fo_ratio.eps}
	    \fi
          }			
	  \caption{\small \label{figure5} Regions of different values of the frameoverlap ratio,
	in $f$ vs $\lambda_i$ space, as obtained from expression (\ref{fo_ratio}). 
	}
    \end{figure}

	Though this needs not necessarily to be the case, in most time-of-flight experiments
	the time window of acquisition is not set to be centered at the elastic line, but instead
	to start when the neutron pulse impinges on the sample, in which case the 
	scattered neutron energies covered extend from very high energies ($\infty$) 
	to a minimum energy value 
	defined by the time window size (which, in this case, would correspond to the maximum neutron
	time-of-flight to be measured).
	In this situation, a generally followed rule of the thumb  
	%when defining the time overlap ratio 
	%to be used in most experiments 
	is that of considering this maximum time-of-flight
	% separating two neutron pulses impinging on the sample as 
	to be given by \cite{Lechner_report1991}
	\begin{equation}
	 t_{\rm max}=1.5\: t_{1{_{\rm el}}}=1.5\sqrt{\frac{m}{2}} \frac{L_1}{\sqrt{E_i}}
	=1.5\frac{m}{h}L_1\lambda_i
	\end{equation}  
	where $t_{1{_{\rm el}}}$ 
	represents the time an elastically scattered neutron (of wavelength $\lambda_i$) takes to 
	travel from the sample to the detector positions (separated by the distance $L_1$). 
	Hence, in these cases, the frameoverlap ratio 
	is determined by the speed of rotation of the chopper system and the wavelength
	of the neutron pulses incident on the sample, as the minimum integer satisfying the condition:
	\begin{eqnarray}
		R\geq t_{\rm max}{_{(\rm s)}}\frac{f_{(\rm rpm)}}{30}=
	%\frac{1.14\times10^{-4}}{\sqrt{E_0{_{(\rm meV)}}}}L_{\rm SD}{_{(\rm m)}}f_{(\rm rpm)}=
	1.26\times 10^{-5}\lambda_i{_{(\rm \AA)}} L_{1}{_{(\rm m)}}f_{(\rm rpm)}
	\label{fo_ratio}
	\end{eqnarray} 		
	%where $f$ represents the frequency of rotation of the disk choppers (in rotations per minute).
	Figure \ref{figure5} presents the regions (in $f$ vs $\lambda_i$ space) 
	corresponding to the different frameoverlap ratios, as calculated from the expression above
	for TOFTOF ($L_1=4$m).

	% third pair of choppers is responsible for the removal of the higher order ......
	%the time spread of the ``chopped'' pulses being shorter the higher is the rotation speed. 

  \subsection{The detector system}

\begin{figure}[H]
      	\centerline
	  {
	    \ifpdf
	    \includegraphics[scale=0.45]{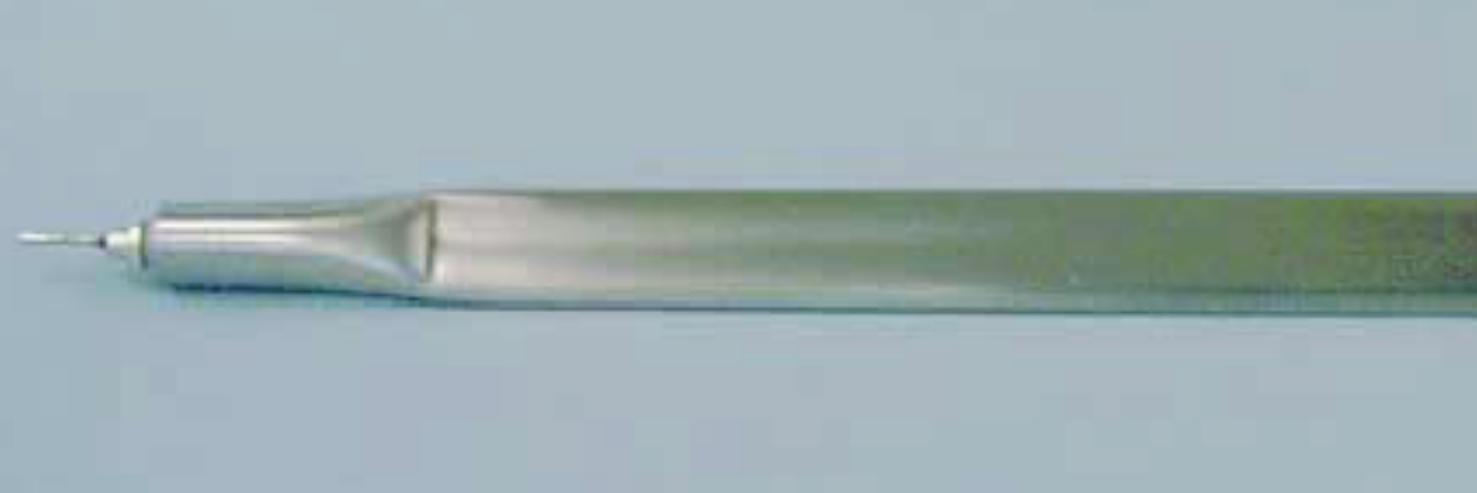}
	    \else
	    \includegraphics[scale=0.45]{toscadet.eps}
	    \fi
	  }			
	  \caption{\small \label{toscadet} Photograph of squashed detectors (from \cite{toscadetectors}).} 
    	\end {figure}

    	TOFTOF's secondary spectrometer contains at the moment a total 605 He$^3$ 
	cylindrical gas detectors, 522 mm long and of 25 mm diameter, squashed 
	to have an almost rectangular section of $15\times30$ mm (figure \ref{toscadet})
	\cite{toscadetectors}.

	They are distributed over eight detector racks covering scattering angles 
	between 7$^\circ$ and 140$^\circ$ (along the equatorial plane and the planes corresponding
	of a vertical component of the scattering angle of 
	-7.8$^\circ$, 7.8$^\circ$ and 15.6$^\circ$ - recall figure 1), and	
	oriented so as to be tangential to surface of a sphere of radius	
	4.000 m centered at the sample position and to Debye-Sherrer circles that,
	in that sphere cover the scattering angles mentioned above.

\begin{figure}[H]
      \centerline
	  {
	    \ifpdf
	   \includegraphics[scale=0.7]{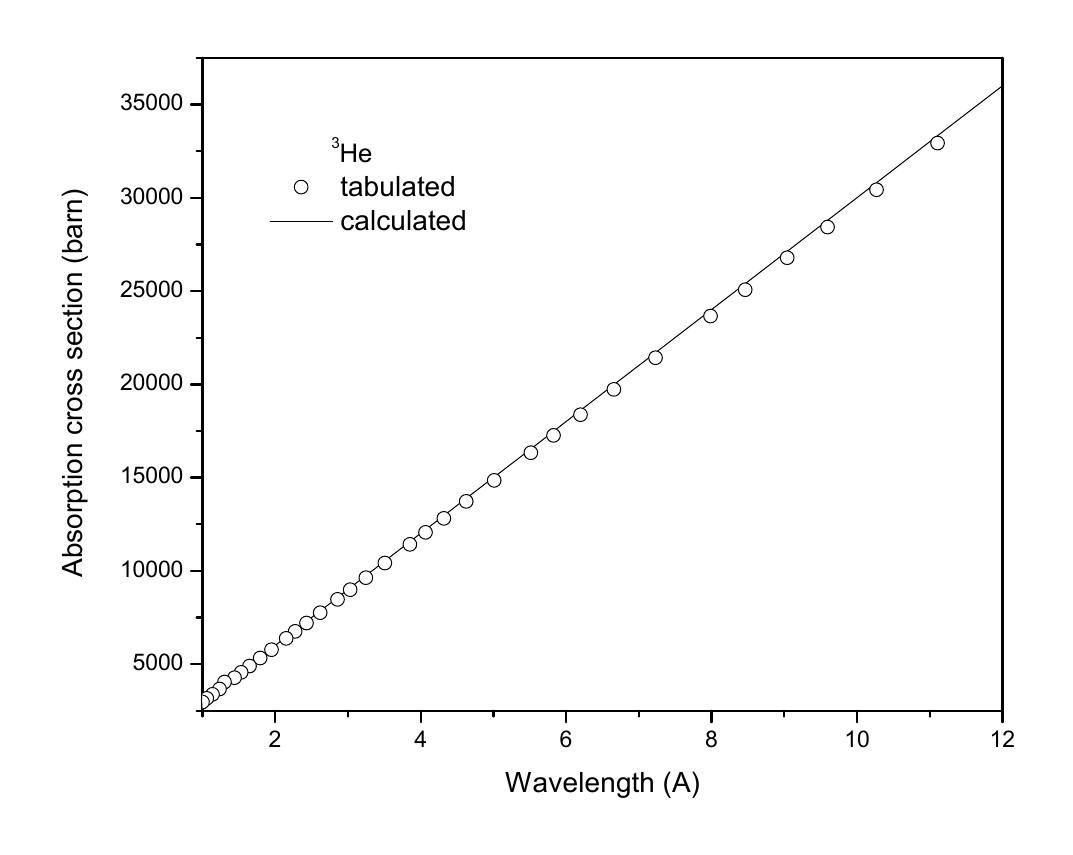}
	    \else
	   \includegraphics[scale=0.7]{Heabscross.eps}
	    \fi
	  }			
	  \caption{\small \label{fig_absHe} Total absorption cross-section for $^3He$,
            as a function of neutron wavelength: scattered pointed tabulated values 
            \protect\cite{nuclear_cross}; full line 
            calculated from $\sigma_a(E)=\sigma_0 \lambda$, with $\sigma_0= 3000$ barn, 
	for $\lambda=1$\AA.} 
    \end{figure}

    There are detector performance parameters, such as the detector
    efficiency or the detector effective thickness , which depend on 
    the energy of the detected neutrons. This results from the fact that 
    that neutrons are detected via absorption in the gas medium of the 
    detector and that the absorption 
    cross-sections vary linearly with the neutron wavelength in the wavelength
    range of interest, as shown in  figure \ref{fig_absHe}. So, unless all the 
    neutrons to be detected have the same energy, the dependence of these parameters 
    has to be properly accounted for, during instrumental performance 
    analysis and data handling.
    
    %The dependence of the absorption cross-sections on the energy of the 
    %detected neutron (figure \ref{fig_absHe})
    %propagates to the some relevant detector performance parameters, such as the 
    %detector 
    %efficiency and the effective detector thickness. So, unless all the 
    %neutrons to be detected have the same energy, this dependence has to be 
    %calculated and properly accounted for during instrumental performance analysis and 
    %data handling.
    The neutron flux across a layer at a given depth {\it x} from the surface of a 
    slab detector is given by: 
    \begin{equation}
      n(x)=n_0e^{-N\sigma_a(\lambda) x}
      \label{detect_abs}
    \end{equation}
    where $n_0$ represents the incident neutron flux  and {\it N} 
    is the density of absorbing atoms with neutron absorption 
    cross-section $\sigma_a(E)=\sigma_0\lambda$ (see figure \ref{fig_absHe}).  
    
    The absorption efficiency of a detector, defined by the fraction of neutrons 
    entering the
    detector, which are actually absorbed by the detector material, is then, in 
    the case of a slab detector of thickness {\it d}, given by: 
    %The efficiency of a detector is defined by the fraction of neutrons 
    %entering the
    %detector, which are actually absorbed by the detector material and result in an 
    %output pulse from the detector. Hence, in the case of a slab detector of thickness 
    %{\it d} it is given by: 
    \begin{equation}
      \eta(d,\lambda)=\frac{n(d)-n(0)}{n(0)}=\left(1-e^{-N\sigma_a(\lambda)d}\right)
      \label{detect_eff}
    \end{equation}
    with $\eta$ depending on the wavelength of the detected neutrons through the wavelength
    dependence of the absorption cross-section of the detector material.

    The detection efficiency is generally defined by the product of this quantity with
    another one, generally designated as neutron sensitivity. The latter represents the 
    fraction of absorption events that result in an electric pulse from the detector.
    In the case of gas detectors, the neutron sensitivity is normally very close to 
    unity, for a proper set up of the detector voltage and discriminator. 
    %where $\varepsilon$, the neutron sensitivity, is the fraction of absorption 
    %events that result in an 
    %electric pulse from the detector, normally very close to unity, for proper
    %set up of the detector voltage and discriminator. 
    
    From equation (\ref{detect_eff}) one can see that the efficiency value can be 
    improved either by increasing the absorbers density
    or by increasing the detector thickness. 
    Because of the first, gas detectors usually contain gas under pressure. 
    For instance, the TOFTOF detectors are operated at 10 atm.
    %The
    %density of absorbing atoms is, nevertheless, limited by pressure constraints
    %to values of the order of $\sim 10^{20}$ cm$^{-3}$
    %\footnote{This value is much smaller than the
    %  ones that can be achieved with the scintillator detectors where solid slabs
    %  of absorbing materials are used and hence much higher
    %  densities of absorbing atoms can be achieved. This actually
    %  constitutes the main advantage of the scintillators over the gas detectors.}.
    % In figure \ref{efficiency}
    %the variation of the efficiency value with neutron energy is represented.
    As for the detector thickness parameter, 
    whereever the time-of-flight technique is used, it affects also  
    the instrument resolution, so any increase in detector thickness must be carefully 
	evaluated. 

    %In fact, to an uncertainty 
    %in the detection position ($\Delta x$) corresponds an uncertainty 
    %in detection time ($\Delta t$), 
    %which contributes to the total neutron time-of-flight uncertainty. 
    %Simply by differentiation of equation (\ref{tflight}) 
    %with respect to the flight-path, the relation between the two can be obtained: 
    %\begin{equation}
    %  \Delta t=\sqrt{\frac{m}{2}}E^{-1/2}\Delta x
    %  \label{deltat_d}
    %\end{equation}
    %In order to have an idea of 
    %the importance of this effect its enough to think that, for a 100 meV neutron, 
    %an uncertainty in detection position of 2.5 mm already contributes with 
    %0.57 $\mu s$ to the total time-of-flight uncertainty. 
    The average detection position $\bar x$
    %(essential for an accurate determination of the flight path length) 
    and an estimation of its uncertainty $\Delta x$ (usually
    referred as {\it effective thickness}, can be obtained from the centroid and root 
    mean square value of the {\it n(x)} distribution, respectively.
    In the case of a slab detector of thickness {\it d}, the integrals: 
    %\begin{eqnarray}
    %         \overline {x}(d,E)=\frac{1}{N\sigma_a}
    %         \frac{1-e^{-N\sigma_ad}(1+N\sigma_ad)}{1-e^{-N\sigma_ad}}\\
    %         \nonumber\\
    %         \Delta {x(d,E)}=\sqrt{8\ln 2}\sqrt{\overline {x^2}(d,E)-\overline {x}^2(d,E)}\qquad\qquad
    % \label{detect_thick}
    %\end{eqnarray}
    \begin{eqnarray}
      \zeta_0(d,\lambda)=\int_0^d n(x)dx \qquad \zeta_1(d,\lambda)=\int_0^d x n(x) dx
      \qquad \zeta_2(d,\lambda)=\int_0^d x^2 n(x) dx 
    \end{eqnarray} 
    can be solved analytically\protect\footnote{In the case of a detector with a circular cross section in the plane of the neutron 
    	beam,
    	each slice of the circle at a distance
    	{\it z} from the center will have a different thickness given
    	by $d(z)=2\sqrt{R^2-z^2}$. In this case, 
    	expressions for the detector
    	parameters can be obtained by proper averaging over {\it z}.\\
    	The detection efficiency is therefore given by:
    	\begin{eqnarray}
      	\eta^*(R)=\frac{1}{R} \int_{0}^R \eta(2\sqrt{R^2-z^2})\; dz \nonumber\\
      	\qquad = \int_{0}^1 1-e^{-N\sigma_a 2R
        \sqrt{1-u^2}} du \nonumber
      	\label{detect_sec}
    	\end{eqnarray}
    	where $u=z/R$ and the integral can only be evaluated numerically.\\
    	Similarly, average detection position and effective thickness can be obtained 
    	numerically from:
    	\begin{eqnarray}
      	\bar {x}\;^*(R)=\frac{\int_{0}^1 \zeta_1(2R\sqrt{1-u^2})\; du}
           {\int_{0}^1 \zeta_0(2R\sqrt{1-u^2})\; du} \qquad\nonumber\\
           \nonumber\\
           % =\frac{1}{N\sigma_a}
           % \frac{\int_{0}^1 1-e^{-N\sigma_a2R\sqrt{1-u^2}}
           %   (1+N\sigma_a2R\sqrt{1-u^2}) du}
           %      {\int_{0}^1 1-e^{-N\sigma_a 2R \sqrt{1-u^2}} du }\\
           %\nonumber\\
           \Delta {x^*(R)}=\sqrt{8\ln 2}\;\sqrt{\overline {x^2}^*(R)-\overline {x^*}^2(R)}
	\nonumber
    	\end{eqnarray}
    	with $\overline {x^2}\;^*(R)=\frac{\int_{0}^1 \zeta_2(2R\sqrt{1-u^2})\; 
      	du}{\int_{0}^1 \zeta_0(2R\sqrt{1-u^2})\; du}$.\\
	Nonetheless, since the TOFTOF detectors are squashed detectors, they are here assumed to 
	the well described as detectors of rectangular shape.
    }
    and hence $\bar x$ and $\Delta x$ are expressed by:
    \begin{eqnarray}
      \overline {x}(d,\lambda)=\frac{\zeta_1(d,\lambda)}{\zeta_0(d,\lambda)}=\frac{1}{N\sigma_a}
      \frac{1-e^{-N\sigma_ad}(1+N\sigma_ad)}{1-e^{-N\sigma_ad}}\\
      \nonumber\\
      \Delta {x(d,\lambda)}=
	\sqrt{8\ln 2}\;\sqrt{\overline {x^2}(d,\lambda)-\overline {x}^2(d,\lambda)}\qquad  
      	%       \nonumber \\
      	%       \Delta {x(d,E)}=\sqrt{\overline {x^2}(d,E)-\overline {x}^2(d,E)}
      	%       \Delta {x(d,E)}=\sqrt{\frac{\zeta_2(d,E)}{\zeta_0(d,E)}-
      	%         {\overline {x}^2(d,E)}}=
      	%       \qquad\qquad\qquad\qquad\qquad\qquad\qquad\qquad\qquad\qquad
      	%       \nonumber \\ 
      	%       =\frac{1}{N\sigma_a}\sqrt{\frac{2(1-e^{-N\sigma_ad}
      	%             (1+N\sigma_ad+\frac{{N\sigma_ad}^2}{2}))}{1-e^{-N\sigma_ad}}-
      	%           \left(\frac{1-e^{-N\sigma_ad}(1+N\sigma_ad)}{1-e^{-N\sigma_ad}}
      	%           \right)^2}
      \label{detect_thick}
    \end{eqnarray}
    with $\overline {x^2}(d,\lambda)=\frac{\zeta_2(d,\lambda)}{\zeta_0(d,\lambda)}=\frac{1}{N^2\sigma_a^2}
    \frac{2(1-e^{-N\sigma_ad}(1+N\sigma_ad+\frac{{N\sigma_ad}^2}{2}))}{1-e^{-N\sigma_ad}}$.\\

\begin{figure}
      \centerline
	  {
	    \ifpdf
	    \includegraphics[scale=1.05]{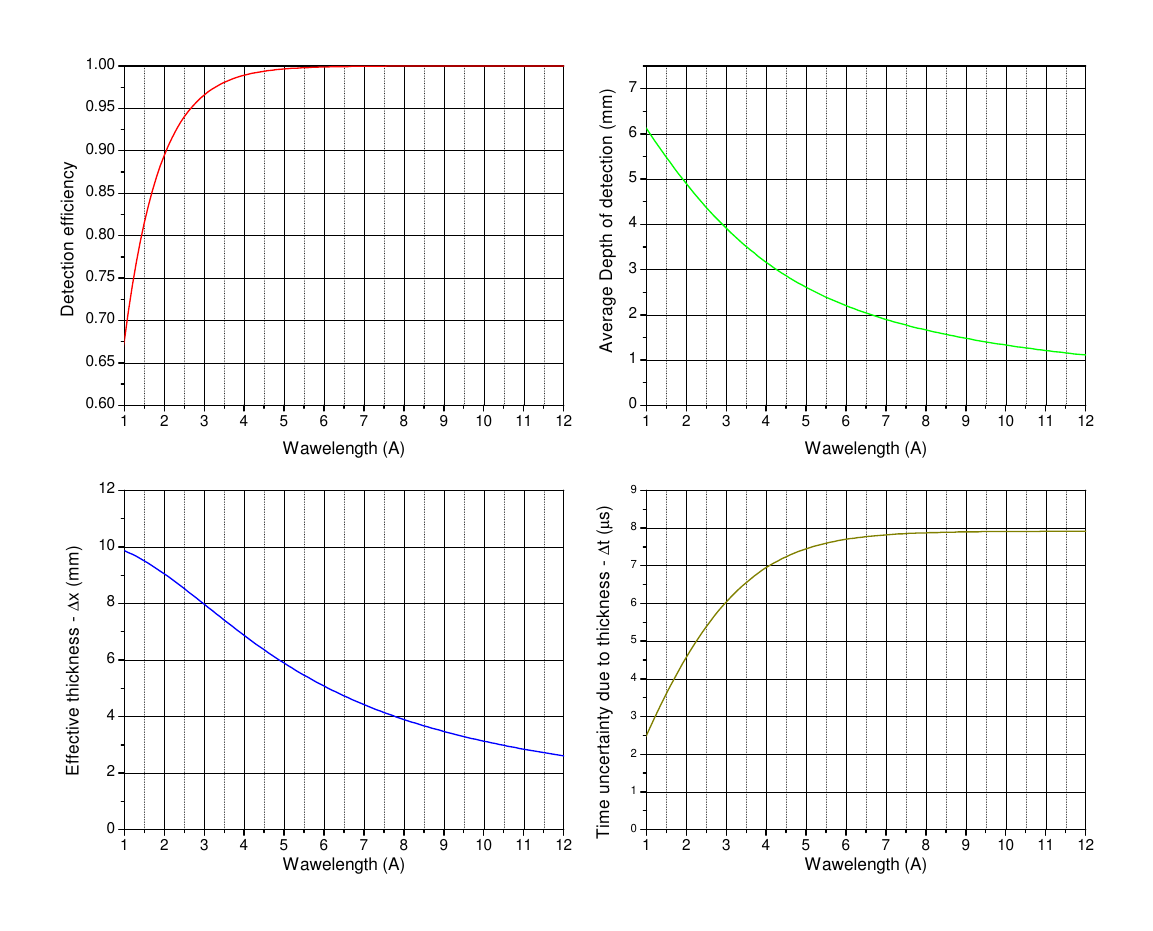}
	    \else
	    \includegraphics[scale=1.05]{TOFTOF_detpar.eps}
	    \fi
	  }			
	  \caption{\small \label{fig_det_par} Detection efficiency, detection average depth, 
            effective thickness and detector thickness contribution to the total time uncertainty
            as a function of neutron wavelength. Calculated for detectors filled with 
            $^3He$ at 10 atm ($N= 2.45\times 10^{20}\; cm^{-3}$ and $\sigma_a=3000\times 10^{-24}\lambda$ 
	cm$^2$, for $\lambda$ in \AA).} 
    \end {figure}

    The repercussion of the absorption cross-section dependence on the neutron wavelength
    on the $\eta$,$\bar x$ and $\Delta x$ functions of TOFTOF detectors is shown 
    in figure \ref{fig_det_par}, 
    %	where some results obtained for both slab and 
    %circular cross-sections are presented, 
    together with the corresponding time-uncertainty, 
    which represents 
    the detector contribution to the total neutron pulse spread:
	\begin{equation}
	   \Delta t =\frac{m\lambda_f}{h}\Delta x
	\label{detector_fwhm}
	\end{equation}
    where $h$ stands for the Planck constant and $m$ for the neutron mass.
   
	 An also important property of a detection system is its {\it dead-time}. 
    This designates 
    the maximum interval of time separating two absorption events that cannot
    be distinguished by the detector. In some cases the limiting time may be set 
    by the processes in the detector itself, and in other cases the limit may arise
    in the associated electronics. In the case of the gas proportional counters it
    is mainly determined by the time involved in the drift of the electrons from a 
    neutron absorption site to the anode. Hence, in these detectors, dead-time 
    depends essentially on the counter cathode and anode diameters 
    and on the chosen anode voltage \cite[chap.4]{egelstaff_n}. 
    %In the case of scintillation detectors the dead time limit arises from the 
    %associated electronics. In either case 
    Typical values are of $\sim$1 - 2 $\mu$s. 
	
	%\subsection{Pulsed neutrons and the time-of-flight technique}

  %\newpage
  \section{Time-of-flight, energy transfer and momentum transfer}

   As already mentioned, 
   at time-of-flight spectrometers, as TOFTOF, neutrons are counted as a function of their total flight time
    and conversion of this time scale to an energy transfer scale is later performed.
  
    The total neutron flight time may be simply expressed by:
    \begin{equation}
      t=t_i+t_f=\frac{L_i}{v_i}+\frac{L_f}{v_f}=
      \sqrt{\frac{m}{2}}\; \left(\frac{L_i}{\sqrt{E_i}}+\frac{L_f}{\sqrt{E_f}}\right)
      \label{ttime}
    \end{equation}
    where $t_i$, $t_f$, $L_i$, $L_f$, $v_i$, $v_f$, $E_i$ and $E_f$ represent the incident and scattered 
    neutron flight times, flight path lengths, velocities and energies, respectively.
    
    On the other hand, energy conservation law gives:
    \begin{equation}
      E=E_i-E_f
      %\label{detect_eff}
    \end{equation}
    for $E$ representing the energy transfered to the scattering system during the scattering
    process (or, by other words, the neutron energy loss).
 
    Since on TOFTOF spectrometer the incident energy is fixed by
    the chopper system, the scattered neutron energy can be substituted in equation (\ref{ttime})
    by $E_f=E_i-E$, giving the time-of-flight to energy transfer conversion expression:
    % \begin{equation}
    %	 t=\sqrt{\frac{2}{m}}\; \left(\frac{L_i}{\sqrt{E_i}}+\frac{L_f}{\sqrt{E_i-E}}\right)
    %	 \label{tehet}
    %	\end{equation}
    \begin{equation}
      E=E_i-{\frac{m}{2}}\;\frac{L_f^2}{\left(t-\sqrt{\frac{m}{2}}\;\frac{L_i}{\sqrt{E_i}}\right)^2}
      \label{ethet}
    \end{equation} 

\begin{figure}
      \centerline
	  {
	    \ifpdf
	    \includegraphics[scale=0.75]{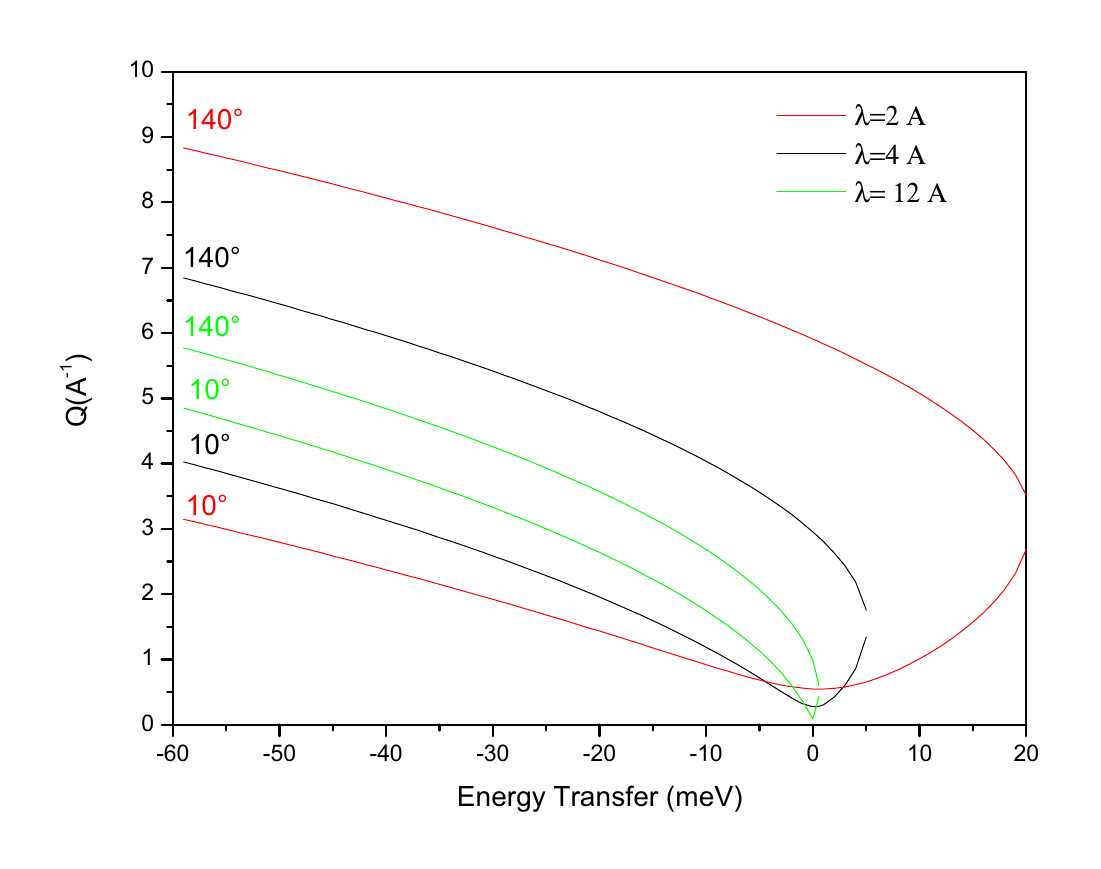}
            \else
	    \includegraphics[scale=0.75]{TOFTOF_Qs1.eps}
	    \fi
          }			
	  \caption{\small \label{toftof_qs1} Trajectories in (Q,E) space scanned with TOFTOF. 
	Variation with scattering angle and incident beam wavelength.
	}
    \end{figure}

\begin{figure}
      \centerline
	  {
	    \ifpdf
	    \includegraphics[scale=0.3]{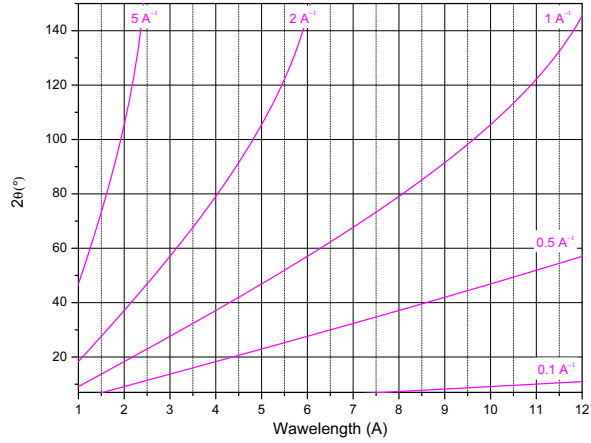}
            \else
	    \includegraphics[scale=0.3]{TOFTOF_Qs.eps}
	    \fi
          }			
	  \caption{\small \label{figqvse} 
          Dependence of the momentum transfered in an elastic scattering process on the neutron incident wavelength
          and neutron scattering angle.
          %Wavelength and angular dependence of the trajectories in Q space scanned with TOFTOF.
          }  
    \end{figure}

    Apart from the energy transfer, the neutron also transfers momentum to the sample, whose amplitude 
    {\it Q}, can be calculated from the conservation law of momentum and the cosine rule: 
    \begin{eqnarray}
      \vec Q=\vec {k_i}-\vec{k_f}	 \qquad \qquad \qquad\nonumber\\
      Q^2=4\pi^2\frac{2m}{\hbar ^2}(E_i+E_f-2 \sqrt{E_i E_f}\cos{2\theta})
      \label{eqlaw}
    \end{eqnarray}
    $2\theta$ being 
    the scattering angle. This equation, together with the conservation law of 
    energy determines the trajectories of a given measurement in the $(Q,E)$
    space.     
    Figure \ref{toftof_qs1} represents the dependence of these trajectories on $2\theta$ 
	and $\lambda_i$
	while, for the particular case to the elastic line, figure \ref{figqvse} represents some of the 
	trajectories corresponding to the same value of $Q$ in $2\theta$ vs $\lambda_i$ space.	    
   %The vectorial representation of the relation expressed by 
    %equation (\protect\ref{eqlaw}) for direct (i) and inverted (ii) spectrometers are displayed in figure
    % \ref{qvect}.
    %The $(Q^2,E)$ trajectories scanned with TOSCA and the average of the ones scanned with the different
    %detector banks of HET are represented in figure \ref{figqvse}. 
    %In the case of time-of-flight observations, 
    %performed at a fixed angular position,
    %to each energy transfer value will correspond a different {\it Q} value.
    
  \section{Instrument resolution}

%  \subsection{Method of calculation}      

     The distances and instants of time relevant for the evaluation of TOFTOF's 
     resolution function are: $L_0$= 10.0 m, the distance between the pulsing and monochromating 
	pairs of choppers, $L_2$= 1.4 m,
	%neutron incidence velocity is defined by a chopper placed at a distance 
	the distance between the monochromating choppers and the sample position, L$_1$=4 m, the distance from the 
	sample position to the detectors positions,
	%there is also a short flight path , which separates
     	%the neutrons scattered by the rotor from the sample area L$_0$, where
	$t_p$, the instant of departure
     of the neutron pulse from the pulsing pair of choppers, 
	and $t_{m}$, $t_s$ and $t_d$, the instants of arrival of the 
     pulse at the monochromating pair of choppers, sample and detector positions, respectively.
     
     %\begin{figure}[H]
     %  \centerline
%	   {
%	     \ifpdf
%	     %\includegraphics[scale=0.325]{hetsqueme.pdf}
%             \else
%	     %\includegraphics[scale=0.325]{hetsqueme.eps}
%	     \fi
%           }			
%	   \caption{\small \label{HETsqueme} Schematic representation of TOFTOF spectrometer}  
%     \end{figure}

     %The distance between the pulsing and monochromating pairs of choppers is 
	%neutron incidence velocity is defined by a chopper placed at a distance 
	%$L_0$= 10.0 m. Between the last choppers and the sample there is also a short flight 
     %path $L_2$= 1.4 m, which separates
     %the neutrons scattered by the rotor from the sample area. 
	%Neutrons transmitted by the 
	%monochromating choppers
     	%drift over this flight path with little dispersion in time and 
     	%arrive at the sample position in a pulse. 
     The arrival time $t_s$ of the pulse at the sample may then be expressed in terms of the
	pior instants of time and distances as
     \begin{equation}
	 t_s=t_p+\frac{L_0+L_2}{v_0}=t_{p}+(t_{m}-t_p)\frac{L_0+L_2}{L_0}=
	\left(1+\frac{L_2}{L_0}\right)t_{m}-
       \left(\frac{L_2}{L_0}\right)t_p
      % t_s=t_p+\frac{L_{_{PM}}L_{_{MS}}}{v_0}=t_{p}+(t_{m}-t_p)\frac{L_{_{PM}}+L_{_{MS}}}{L_{_{MS}}}=
	%\left(1+\frac{L_{_{MS}}}{L_{_{PM}}}\right)t_{m}-
       %\left(\frac{L_{_{MS}}}{L_{_{PM}}}\right)t_p
       %\label{xx}
     \end{equation}

     After the scattering process the neutrons travel a distance $L_1=4.0$ m, 
     from the sample to the detector, the 
     scattered neutron energy being determined by its flight-time $t_f=t_d-t_s$. 
	Expression (\ref{ethet})
     can then be re-written as:
     \begin{eqnarray}
       E=\frac{1}{2}m\left\{\frac{L_0^2}{(t_m-t_p)^2}
       -\frac{L_1^2}{\left[t_d-\left(1+\frac{L_2}{L_0}\right)t_{m}+
           \left(\frac{L_2}{L_0}\right)t_p\right]^2}\right\}
       \label{ethett}
     \end{eqnarray} 
     and the uncertainty $\Delta E$ can be obtained within a reasonable approximation
	from
     \begin{equation}
      	\Delta E=\sqrt{\sum_x\left(\frac{\partial E}{\partial t_x}\Delta t_x\right)^2}
      	\label{resres}
     \end{equation}
    where $\Delta t_x$ represents the uncertainties associated with each of the time instants.
	Hence, one obtains\footnote{Note that this expression is equivalent to that presented R E Lechner \cite{Lechner_report1991}.}:
     \begin{equation}
       \footnotesize
       \Delta E=2\sqrt{\frac{2}{m}}\;\sqrt{
         \left(\frac{E_i^{\frac{3}{2}}}{L_0}+
	\frac{L_2}{L_0}\;\frac{E_f^{\frac{3}{2}}}{L_1}\right)^2(\Delta t_p)^2+ 
         \left(\frac{E_i^{\frac{3}{2}}}{L_0}+\left(1+\frac{L_2}{L_0}\right)
	\frac{E_f^{\frac{3}{2}}}{L_1}\right)^2
         (\Delta t_{m})^2+
         \left(\frac{E_f^{\frac{3}{2}}}{L_1}\;\Delta t_d\right)^2}    
       \label{hetdeltae}
     \end{equation} 
     with $E_f=E_i-E$.
     The uncertainties $\Delta t_p$ and $\Delta t_m$ are given by the opening times of the 
	pulsing and monochromating pairs of choppers
	%obtained from expression (\ref{chopper_fwhm}) 
	and $\Delta t_d$ is obtained from
     the convolution of all the time uncertainties associated with the performance of the 
	detectors (dead-times, uncertainty in detection position) with the uncertainties in the 
	scattered neutrons flight-path. Here,
     $\Delta t_d$ was considered to be given by the convolution of a detection dead-time of 1$\mu$s
     with the time-uncertainty associated with the uncertainty in detection depth 
	(expression (\ref{detector_fwhm})) and the time uncertainties resultant from 
	the other two coordinates representing the 
     detection position (due to the effective length and diameter of the detectors)\footnote{ 
       These quantities may be estimated from:
       \begin{equation}
         \Delta t = \sqrt{\frac{m}{2}}\frac{\Delta L_1}{\sqrt{E_f}}\qquad \qquad 
         \Delta L_1= L_1\left (\frac{1}{\cos(\tan^{-1} {\frac{x}{2 L_1}})}-1\right)\nonumber
       \end{equation}
       where $E_f=E-E_i$, $x$ represents either the effective length (400 mm) 
       of the effective width of the detector (30 mm),
       and $L_1=4$ m. 
       In the calculations presented the uncertainty due to the finite detector
       width was disregarded and only the uncertainty due to the detector length was considered, since the 
       former was one order of magnitude smaller than the latter.
     }.
   
     Expression (\ref{hetdeltae}) can be written in a more intuitive way as:
     \begin{equation}
       %\footnotesize
       \Delta E=2\sqrt{\frac{2}{m}}\;\frac{E_f^{\frac{3}{2}}}{L_1}\;\Delta t
       \label{hetdeltae2}
     \end{equation} 
     where $\Delta t$ is given by:
     \begin{equation}
       \footnotesize
       \Delta t=\sqrt{
         \left[\left(\frac{E_i}{E_f}\right)^{\frac{3}{2}}\frac{L_1}{L_0}+\frac{L_2}{L_0}\right]^2
	(\Delta t_p)^2+ 
         \left[\left(\frac{E_i}{E_f}\right)^{\frac{3}{2}}\frac{L_1}{L_0} +
	\left(1+\frac{L_2}{L_0}\right)\right]^2
         (\Delta t_{m})^2+
         (\Delta t_d)^2}  
     \end{equation}

     %This way of expressing the instrument resolution function highlights the role-played by the
     %distance L$_1$ in the definition the instrument� resolution curve. $\Delta t$ 	
     
     %\begin{figure}
     %  \centerline
%	   {
%	     \ifpdf
%	     %\includegraphics[scale=0.4]{het_res2_04.pdf}
%             \else
%	     %\includegraphics[scale=0.4]{het_res2_04.eps}
%	     \fi
%           }			
%	   \caption{\small \label{res2_HET} Energy uncertainty $\Delta E $ and corresponding 
%             resolution curve $\Delta E/E$
%         of HET spectrometer as a function of incident neutron energy $E_i$ (chopper B, 500Hz, $L_f=$4 m).}  
%     \end{figure}
    \begin{figure}
       \centerline
	   {
	     \ifpdf
	     \includegraphics[scale=0.6]{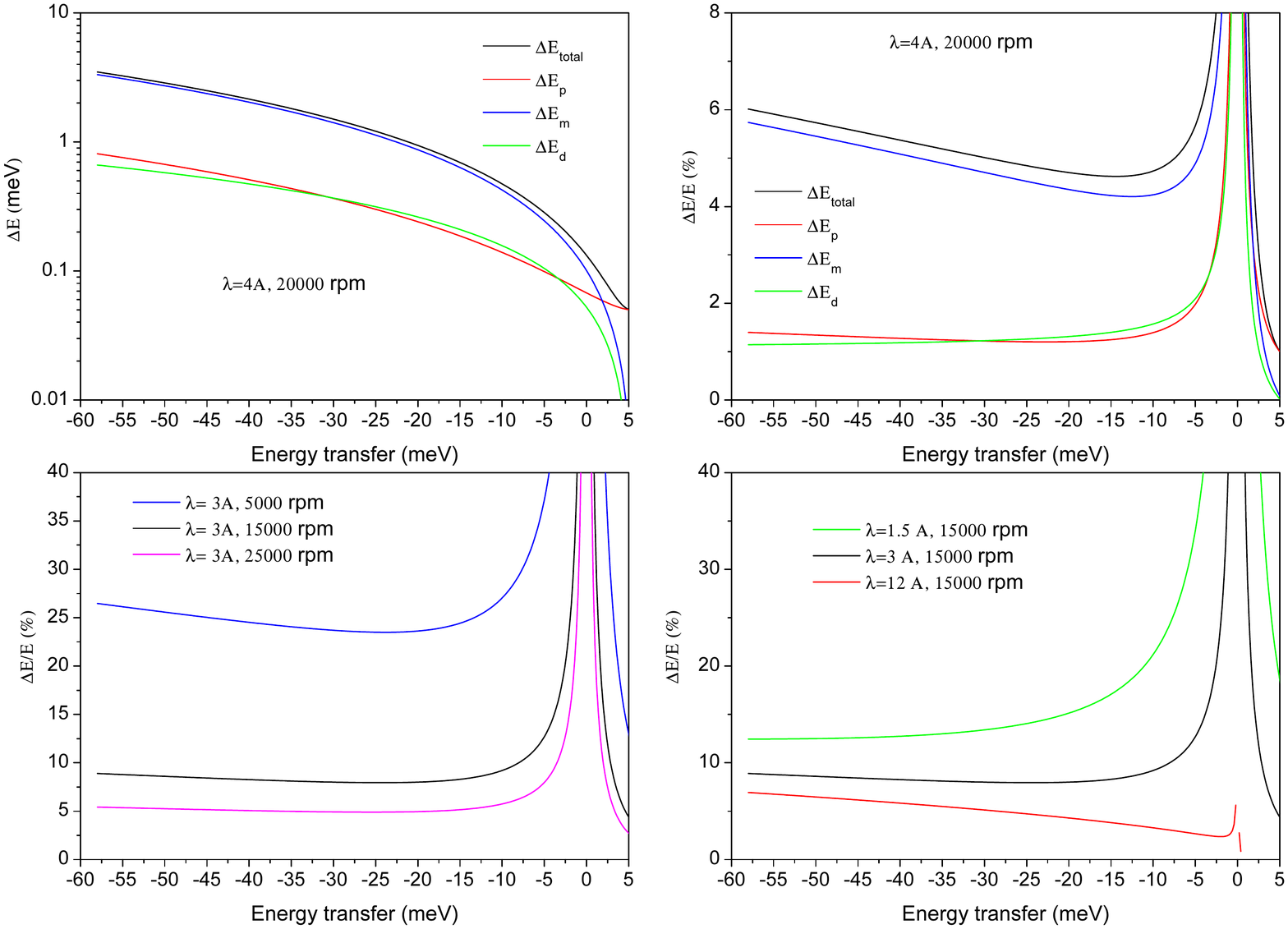}
             \else
	     \includegraphics[scale=0.6]{TOFTOF_resresults.eps}
	     \fi
           }			
	   \caption{\small \label{res_HET} 
	TOP: Uncertainty in energy transfer and resolution of TOFTOF spectrometer for
             an incident neutron wavelength of $4 \AA$ (and a choppers speed of 20000 rpm):total and partial
	contributions.
	BOTTOM: Variation of the instrument resolution with wavelength and choppers speed. Calculations
	performed considering the use of the larger set of slits. 
	 }  
     \end{figure}

     In figure \ref{res_HET}, the uncertainty $\Delta E$ 
	associated with each value of 
     the neutron energy transfer $E$
     is represented for some particular choices of neutron chopper velocity and incident energy,
	together with the corresponding resolution function $\Delta E/E$. 
     For one set of parameters ($\lambda$= 4 \AA, choppers speed 20000 rpm, 
	large slits for both monochromating and pulsing pairs of choppers) 
	the individual contributions 
     are also displayed.One can seen that the major contributor to the instrument
	resolution is the monochromating CR pair of choppers. The contribution from the pulsing pair 
	of choppers only dominates in the region of neutrons energy loss.  	
	The contribution of the 
	secondary part of the spectrometer %to the instrument resolution
	is generally smaller than that of the primary part of the spectrometer (when using 
	large slits), except 
	for situations of both high wavelengths and high chopper speeds, in the region around the 
	elastic line.
   	
	The use of the smaller slits at the pulsing and monochromating pairs of choppers, would reduce
	the contribution of the primary part of the spectrometer to half its value while keeping 
	constant the contribution from the secondary part of the spectrometer. 
	It should be noted, however, that an almost equal resolution improvement can be obtained
	(at the elastic line and over the entire energy gain region)
	by keeping the large slits in use at the pulsing pair of choppers and only changing for the 
	smaller slits at the monochromating pair of choppers. 
 
     %The variation of both $\Delta E$ and the corresponding resolution function $\Delta E/E$ with the
     %	incident neutron wavelength and choppers speed are displayed in figure \ref{res2_HET}.
     
     %It can be seen from the results presented that, at TOFTOF, the uncertainty in energy transfer decreases 
     %as the energy transfer increases, reaching its minimum values for energy loss values close to the 
     %value of the incident energy ($E\sim E_i$), as expected from expression \ref{hetdeltae}.
     %The main contributions to the uncertainty in 
     %energy transfer come from the chopper system. 
     %The moderator term dominates the uncertainty for 
     %energy transfer values approaching those of the incident energy, while the chopper term determines the 
     %uncertainty at the lower values of energy transfer.
     %This explains the observed reduction of the uncertainty in energy transfer at $E\sim E_i$ for increasing 
     %values of $E_i$, above 300 meV (recall figure \ref{fig_deltatmTOSCA}) and well as the increase in 
     %the uncertainties at $E\sim 0$.
     %As expected the uncertainty decreases with increasing chopper velocities and with
     %decreasing values of the chopper slit width (see also table \ref{table_choppers}).
     %This is well visible, particularly in the region of energy transfer where this chopper 
     %contribution exceeds that of the moderator term. 
     \begin{figure}
	\centerline
	   {
	     \ifpdf
	     \includegraphics[scale=0.3]{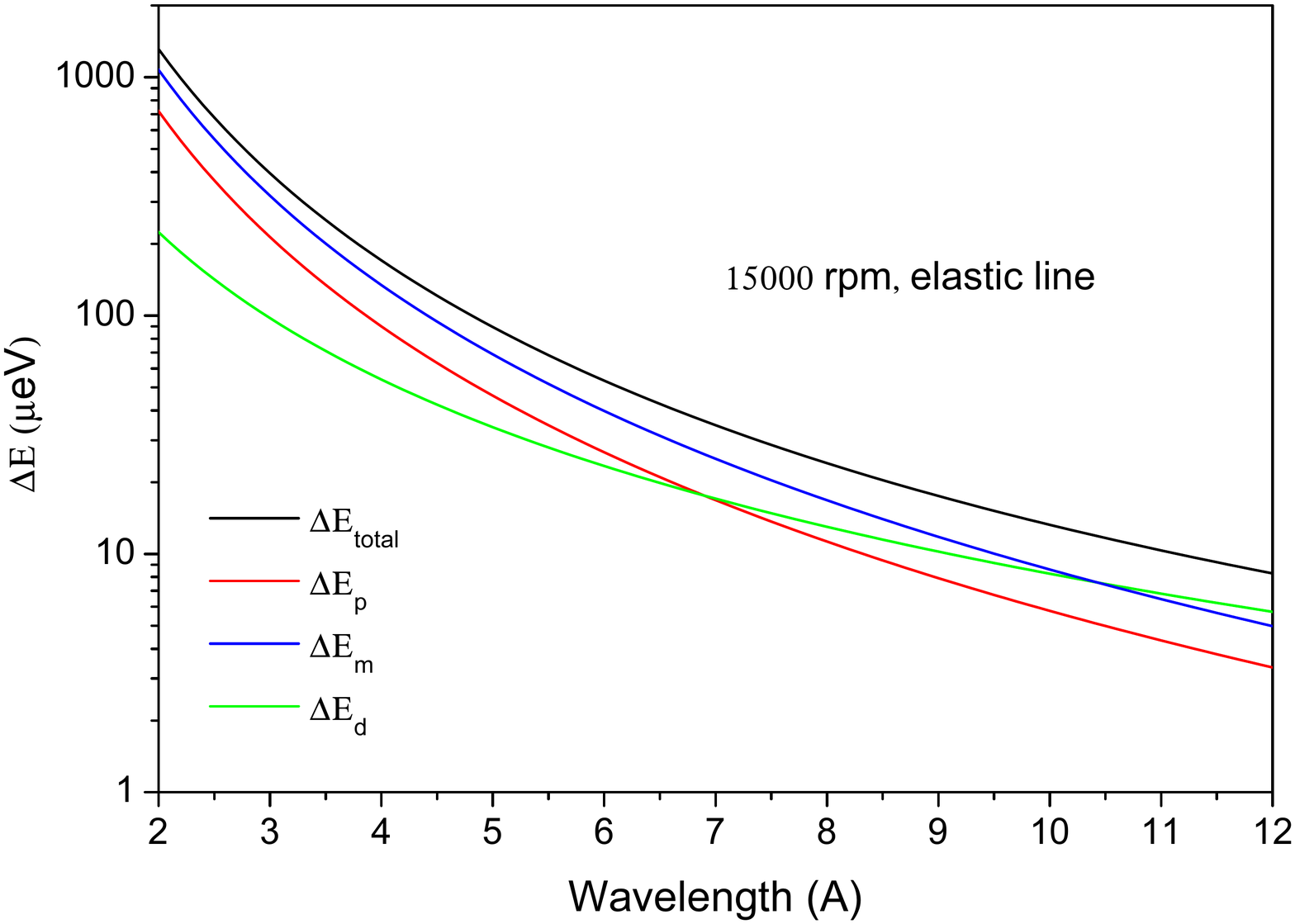}
	     \includegraphics[scale=0.3]{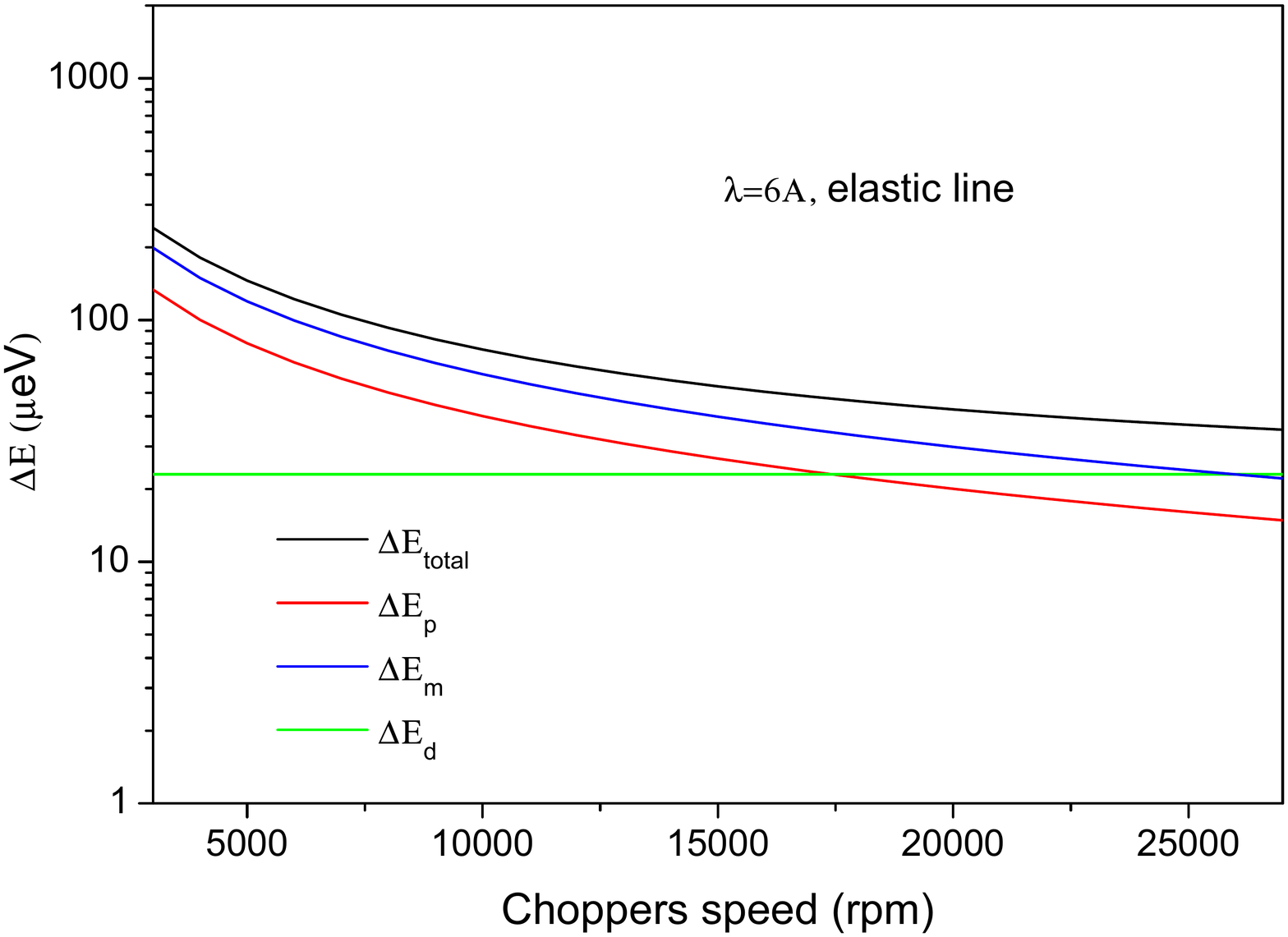}
             \else
	     \includegraphics[scale=0.3]{Res_elcomp1.eps}
	     \includegraphics[scale=0.3]{Res_elcomp2.eps}
	     \fi
           }			
	   \caption{\small \label{figure_n1} Dependence of the energy uncertainty at the elastic line
	on the neutrons wavelength and choppers speed: total and partial contributions.
	%Isoresolution curves at the elastic line
	%in choppers speed vs wavelength space. 
	Calculations performed considering the use of the large slits.}  
     	\end{figure} 

\begin{figure}[H]
       	\centerline
	   {
	     \ifpdf
	     \includegraphics[scale=0.35]{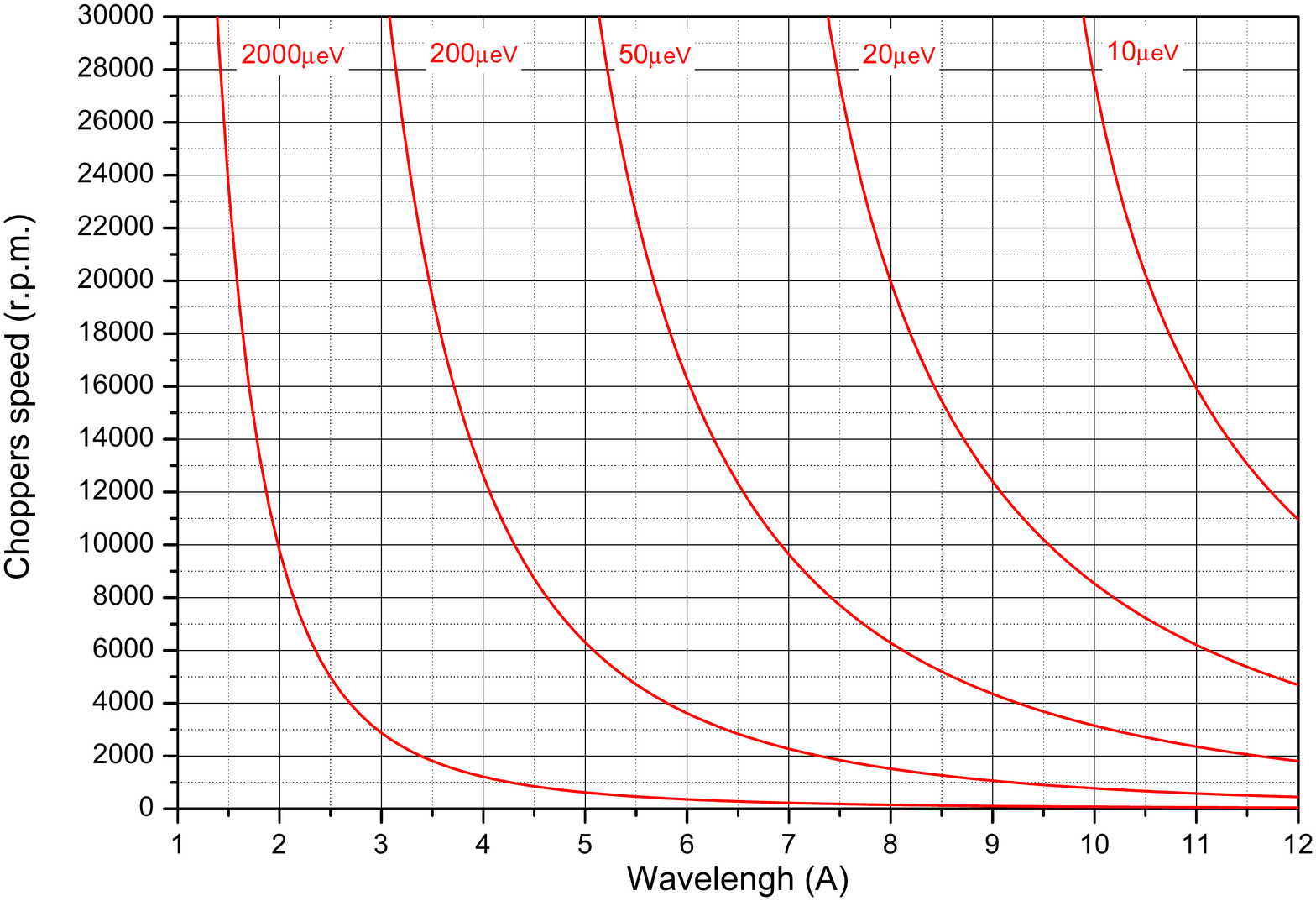}
             \else
	     \includegraphics[scale=0.35]{TOFTOF_isoresolution.eps}
	     \fi
           }			
	   \caption{\small \label{figure_n2} Isoresolution curves at the elastic line
	in choppers speed vs wavelength space. Calculations performed considering the use
	of the large slits.}  
     	\end{figure}

     Specifically with regards to the expected FWHM of the elastic line, the results are summarized in 
     figure \ref{figure_n1} and \ref{figure_n2}, once again for the option of using the largest chopper slits. 
	In Figure \ref{figure_n1} the dependence of the resolution 
	at the elastic line, and its partial contributions, on both the wavelength and the 
	choppers speed is presented. One can observe the increasing importance of 
	the contribution from the secondary part of the 
	spectrometer to the resolution at the elastic line, as both wavelength and choppers 
	speed increase.
        In Figure \ref{figure_n2} some of the isoresolution curves 
	in choppers speed vs wavelength space are represented. 

%\subsection{Quantitative comparison with experimental values}
% 
%  In figure \ref{figure_resvsestim} the resolution estimations described above are compared
%     experimental values, obtained from the determination of the 
%     FWHM of the elastic incoherent scattering peak of a vanadium sample.
%     The agreement is very good. It is noted, however, that 	
%     \begin{figure}
%	\centerline
%	   {
%	     \ifpdf
%	     \includegraphics[scale=0.3]{Resvsestim.pdf}
%             \else
%	     \includegraphics[scale=0.3]{Resvsestim.eps}
%	     \fi
%           }			
%	   \caption{\small \label{figure_resvsestim} Dependence of the energy uncertainty at the elastic line
%	on the neutrons wavelength and choppers speed: total and partial contributions.
%	%Isoresolution curves at the elastic line
%	%in choppers speed vs wavelength space. 
%	Calculations performed considering the use of the large slits.}  
%
 \vskip 12pt

     It will be interesting to extensively compare the results of these predictions with 
     experimental values, obtained from the determination of the FWHM of the elastic incoherent scattering peak of a vanadium 
     sample. 
     Preliminary results, based on the limited experimental data available up to now, indicate
     an excellent agreement.      	
     %Finally, it should be mentioned that the 
     %evolution of the different contributions with energy transfer agree with previous predictions
     %made for HET \cite{Perring,chop} and for 
     %other direct geometry spectrometers (see e.g. \cite{windsor}).
     %The predictions here presented 
     %are also in reasonable quantitative agreement with the experimental measurements 
     %of the elastic incoherent scattering peak FWHM \cite{Perring}.  
     In any case it should be noted that discrepancies of the order of 10\% between the
     values calculated here and those determined experimentally are to be expected, given 
	the approximations involved in the calculations.
     %	 should be considered when dealing 
     %with these predictions. Discrepancies of 10 to 20 \% have also been identified in the previous
     %	 predictions 
     %\cite{Perring,chop}. 
  %\section{Frame overlap ratio}

\begin{figure}[H]
       \centerline
	   {
	     \ifpdf
	     \includegraphics[scale=0.3]{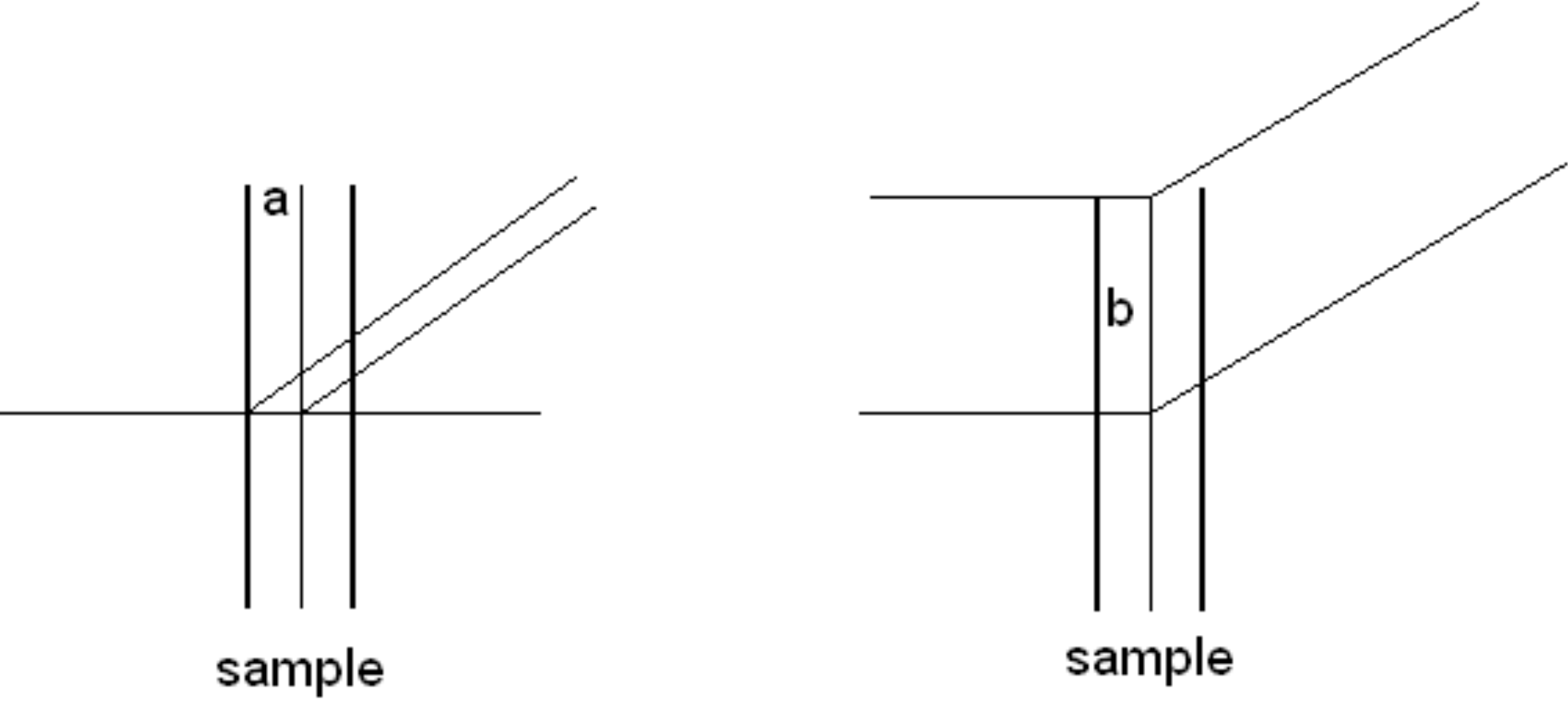}
             \else
	     \includegraphics[scale=0.3]{sample.eps}
	     \fi
           }			
	   \caption{\small \label{sample} The sample dimensions give rise to flight time spread. }  
     \end{figure}

	Another aspect that should be taken into account, when determining the resolution function 
     corresponding to a specific measurement, is the sample dimensions, 
	since  they introduce uncertainties in the neutron
     flight-path. The corresponding time-spread will depend not only on the sample geometry, but 
     also on the incident and scattered neutron energies, as well as on the scattering angle ($2\theta$).  
    For a rectangular sample of width {\it 2c} in the scattering plane and thickness {\it 2a}, 
     %and considering 
     %an orientation of the detector banks perpendicular to the direction of the scattered beam as in the 
     %case of the position sensitive detector bank at HET,
     it will be given by: (figure \ref{sample}):
     \begin{equation}
       \Delta t_s=\sqrt{\frac{m}{2}}\;\sqrt{c^2\frac{(\sin{2\theta})^2}{E_f}+a^2\left(\frac{1}{E_i}+
         \frac{1}{E_f}(\cos{2\theta})^2-\frac{2\cos{2\theta}}{\sqrt{E_i\;E_f}}\right)}
       %\label{xx}
     \end{equation}
     {\it c} being typically a couple of centimeters, because of the neutron beam dimensions, and {\it a}
     being usually set to a value such that the multiple scattering contribution to the signal detected 
     can be neglected (typically varying from some tenths of millimeter to some millimeters). 

     On TOFTOF spectrometer, 
	%the time-spreads introduced by a sample thickness up to some
	%milimetrers may be considered negligible, in face 
	%of the other time uncertainties determining the instrument resolution, except for
     	%energy loss values very close to the incident energy. 
	a sample with the geometry and  
	%lateral 
	dimensions mentioned above 
	%in the scattering plane 
	contributes to worsen the instrument resolution, at the elastic line, 
	by at most 1\% of the neutron incident energy.  
        
        This is not the case, however, for samples of cylindrical geometry for which an angle dependent resolution worsening is 
        introduced, which may be particularly significant at high angles. 
  
	%.

    % for the cases for which {\it b} has to be taken into account. 
    % Hence, it can be concluded that the dimensions of the sample containers used 
    % did not worsen significantly the resolution of the spectrometers.

%\newpage	
  \section{Instrument intensity}

	The intensity of an experiment is generally defined by the number of neutrons arriving 
	at the sample position per unit time.

	At TOFTOF, this quantity can be obtained  multiplying the total number of neutrons 
	transmitted through the chopper system with each pulse by
	the number of pulses per second arriving at the sample position $n_p$.
	The latter quantity 	
	is simply determined by the choppers speed and the frameoverlap ratio through
	\begin{equation}
		n_p=\frac{f_{(\rm rpm)}}{30}\frac{1}{R}
	\label{eq_np}
	\end{equation}
	with $R$ representing the frameoverlap ratio chosen.

	The number of neutrons per unit wavelength 
	transmitted each time a pulse is generated by the pulsing pair of choppers will
	be proportional to the neutron flux per unit wavelength arriving 
	at their entrance, $d\phi/d\lambda_i$(in n/cm$^2$/\AA/s), 
	multiplied by the area of the choppers slits and by the choppers
	opening time $\Delta t_p$. 
	The number of neutrons per pulse further transmitted by the monochromating
	pair of chopper will be then proportional to the ratio of the area of the 
	choppers slits to the area of the neutron guide at their entrance\footnote{In 
          case of a parallel neutron beam, the area ratio to be considered should be 
	that of the area of the monochromating choppers slits to the pulsing chopper slits.
	},
	%\footnote{
	%	Note that given the neutron guide system that preceeds the pulsing choppers as
	%well as the converging neutron guide between the pulsing and the monchromating 
	%pairs of choppers, at TOFTOF, the dependence of the 
	%neutron intensity (n/s) and neutron flux (n/cm$^2$/s)
	%at the sample position, on the widths
	%of the slits of the monochromating pairs of choppers, is the same.
	%}, 
	multiplied by the wavelength band selected during the choppers opening time 
	$\Delta \lambda_i=\frac{h}{mL_0}\Delta t_m$.     
	%The intensity is then, finally, obtained simply by multiplying this number by
	%the number of pulses per second arriving at the sample position $n_p$.	
	
	%This quantity 
	%is proportional to the number of neutrons arriving per unit area and 
	%time at the entrance of the pulsing pair of choppers,to the number of 
	%pulses generated per second $n_p$ by this choppers and
	% to the transmission function of both this 
	%and the monochromating counter-rotating pairs of choppers \footnote{
	%	Note that given the neutron guide system that preceeds the pulsing choppers as
	%well as the converging neutron guide between the pulsing and the monchromating 
	%pairs of choppers, at TOFTOF, the dependence of the 
	%neutron intensity (n/s) and neutron flux (n/cm$^2$/s)
	%at the sample position, on the widths
	%of the slits of the monochromating pairs of choppers, is the same.
	%}.    

	%This is proportional to the number of 
	%pulses generated per second $n_p$, to the fraction of 
	%time during which the pulsing choppers are opened $\Delta t_p$,
	%to the neutron flux per 
	%unit energy $d\phi/d\lambda_0$ feeding the instrument 
	%and to the wavelength bandwidth selected by the monochromating choppers 
	%$\Delta \lambda_i=\frac{h}{mL_0}\Delta t_m$.
	 
	One may then write the following proportionality relation:
	\begin{equation}
	   I\propto n_p b_p\Delta t_p b_m\Delta t_m
	\frac{d\phi}{d\lambda_i}%_{(n/cm$^2$/$\lambda$/s}
	\label{eq_int1}
	\end{equation}
	where $\Delta t_m$ and $\Delta t_p$ are given by expression (\ref{chopper_fwhm})
	and $b_p$ and $b_m$ represent the widths of the pulsing and the monochromating 
	chopper slits. Note also that, for purposes of clarity, the dependence of the 
	instrument intensity on the quantities that are fixed at TOFTOF, such as $L_0$,
	the heights of the chopper slits or the 
	neutron guide dimensions, was not explicitly included in the expression above.

	%$n_p$, the number of pulses arriving per second at the sample position, 
	%is determined by the choppers speed and the frameoverlap ratio through
	%\begin{equation}
	%	n_p=\frac{f_{(\rm rpm)}}{30}\frac{1}{R}
	%\label{eq_np}
	%\end{equation}
	%with $R$ representing the frameoverlap ratio.
	
	Further substitution of expressions (\ref{chopper_fwhm}) and (\ref{eq_np}) 
	in expression (\ref{eq_int1}), gives
	\begin{equation}
		I\propto {b_p^2}{b_m^2}f^{-1}R^{-1}\frac{d\phi}{d\lambda_i}
	\end{equation} 
 	which expresses the instrument intensity dependence on the changeable 
	instrument parameters, specifically, the widths of the pulsing and monochromating
	choppers slits $b_p$ and $b_m$, the choppers system speed $f$, the wavelength chosen 
	for the incident beam (through $d\phi/d\lambda_i$) and the frame overlap
	ratio chosen $R$.
	
\begin{figure}[H]
       	\centerline
	   {
	     \ifpdf
	     \includegraphics[scale=0.35]{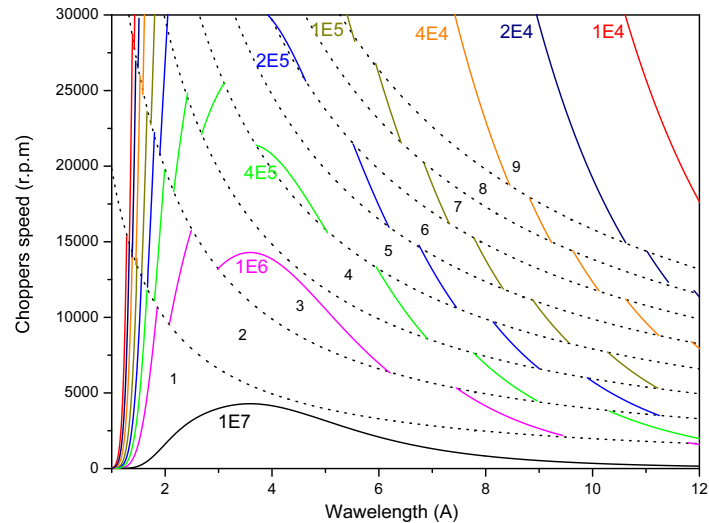}
             \else
	     \includegraphics[scale=0.35]{TOFTOF_intensity.eps}
	     \fi
           }			
	   \caption{\small \label{figure_n} Estimated elastic line isointensity curves in
	wavelength vs choppers speed space, considering the regions of different frameoverlap
	ratio values as defined by expression \ref{fo_ratio} and the use
	of the larger slits of both pulsing and monochromating pairs of choppers.}  
     	\end{figure}

	Figure \ref{figure_n} represents some of the isointensity curves in wavelength versus
	choppers speed space, as calculated considering the regions of different frameoverlap
	ratio defined by expression (\ref{fo_ratio}) and the use of the larger slits in both
	pulsing and monochromating pairs of choppers. 
        Note however that the numerical values displayed result from rough estimations (made 
        on grounds not here discussed: the proportionality constant in expression (24)) and
        hence should be regarded as simply allowing to better understand the dependencies of the 
        instrument intensity on the relevant instrument parameters, i.e. the wavelength of the incident
        neutrons, the choppers speed of rotation and the frame overlap ratio.

	%\paragraph{}
	Note also that 
        %the importance of 
        the secondary part of the spectrometer 
        %to the statistical quality of an actual measurement is not 
        was not taken into account  
        %in the conventional determination of instrumental intensity performed above. 
	%This attitude is probably related to the fact that 
        This was because      
        the number of neutrons per second 
	detected at the instrument detectors will depend on
	the sample being used, while the number of neutrons arriving 
	at the sample position per unit time is uniquely determined by instrument parameters.
	In addition, integration of the secondary part of the spectrometer necessarily introduces 
	an additional degree of freedom, the scattered neutrons energy.	
	Nevertheless, for the purposes of the evaluating of the statistical differences 
	between acquisitions performed under different conditions at the same instrument, one
	should at least include the differences in the detectors efficiency for neutrons of different energies.
	Also when a comparison of the performance of two instruments is intended, it would
	probably be more adequate evaluating 
        integrated intensity of the elastic line, detected 
	at each of the instruments, with the same sample and for the same 
	incident wavelength and resolution configuration. 
	This would take into account not only the primary part of the 
	spectrometers but also the secondary part of the spectrometers, namely the detector angular 
	coverage. 

\section*{Final Note}
To the advantage of the interested reader, the reference list of this manuscript is here extended to include
additional documentation that, being related to the topics treated, was not explicitly cited in 
the text above \cite{Lechner1992,Lechner1996,Schober2000,Ollivier2002,Ollivier2004,Copley2003a,Copley2003b}.

\section*{Acknowledgments}
This work was carried out under a short-term post-doc contract offered by the FRM II and further supported by 
a post-doc research grant (SFRH/BDP/17571/2004) attributed by Fundacao para a Ciencia e Tecnologia. 
%Many other collegues at the FRM II contributed in many different forms. 
%Specifically t
The author greatfully thanks the continuous support and stimulus of the FRM II scientific director, Prof Winfried Petry. 
Dr Tobias Unruh, the TOFTOF's instrument scientist, is also acknowledged for finally motivating the publication of this study. 
	  
%\newpage		

%%%%%%%%%%%%%%%%%%%%%%%%%%%%%%%%%%%%%%%%%%%%%%%%%%%%%%%%%%%%% 
  %***********************************************
%\begin{small}
%  \singlespacing
%  \cleardoublepage
%  \addcontentsline{toc}{section}{References}
%\bf{References\\}
  \bibliography{toftof_arXiv}

\begin{thebibliography}{10}

\bibitem{Zirkel_2000}
A.~Zirkel, S.~Roth, W.~Schneider, J.~Neuhaus, and W.~Petry,
\newblock {\em The time-of-flight spectrometer with cold neutrons at the
  FRM-II},
\newblock Physica B {\bf 276-278} (2000).

\bibitem{Zirkel_report}
A.~Zirkel,
\newblock Zwischenbericht uber den Bau des Flugzeitspektrometer mit kalten
  Neutronen am FRMII,
\newblock Technical report, FRMII - Technische Universitat Muenchen, 2001.

\bibitem{Roth_thesis}
S.~V. Roth,
\newblock {\em Konzeption und neutronenoptische Optimierung des kalten
  Flugzeitspektrometers am FRMII},
\newblock PhD thesis, Technische Universitat Muenchen, 2001.

\bibitem{toftof_webpage}
TOFTOF webpage,
\newblock \url{http://www.frm2.tum.de/toftof/index_en.shtml}.

\bibitem{frm2_webpage}
FRM II webpage,
\newblock \url{http://www.frm2.tum.de}.

\bibitem{Roth_2000}
S.~Roth, A.~Zirkel, J.~Neuhaus, W.~Schneider, and W.~Petry,
\newblock {\em Optimization of the neutron guide system for the time-of-flight
  spectrometer at the FRM-II},
\newblock Physica B {\bf 283} (2000).

\bibitem{TOFTOF_report2004}
T.~Unruh, J.~Ringe, J.~Dörbecker, R.~Funer, A.~Gaspar, J.~Neuhaus, and
  W.~Petry,
\newblock First commisioning steps at the time-of-flight spectrometer TOF-TOF
  at FRMII,
\newblock Technical report, FRMII - Technische Universitat Muenchen, 2004.

\bibitem{Copley1988}
J.~R.~D. Copley,
\newblock {\em On the use of multi-slot multiple disk chopper assemblies to
  pulse thermal neutron beams},
\newblock Nucl Instr and Meth Phys Res A {\bf 273}, 67--76 (1988).

\bibitem{Copley1991}
J.~R.~D. Copley,
\newblock {\em Transmission properties of a counter-rotating pair of disk
  choppers},
\newblock Nucl Instr and Meth Phys Res A {\bf 303}, 332--341 (1991).

\bibitem{windsor}
C.~G. Windsor,
\newblock {\em Pulsed Neutron Scattering},
\newblock Taylor and Francis, 1981.

\bibitem{Lechner_report1991}
R.~E. Lechner,
\newblock TOF-TOF spectrometers at pulsed neutron sources and at steady-state
  reactors,
\newblock Technical report, KEK reports 90-25, 1991.

\bibitem{toscadetectors}
A.~C. inc.,
\newblock Helium-3 Squashed Neutron Detectors,
\newblock \url{http://www.canberra.com/products/1139.asp}.

\bibitem{nuclear_cross}
MCNP and ENDF nuclear data libraries,
\newblock \url{http://atom.kaeri.re.kr/endfplot.shtml} \rm.

\bibitem{egelstaff_n}
P.~A. Egelstaff, editor,
\newblock {\em Thermal neutron scattering},
\newblock Academic Press Inc., 1965.

\bibitem{Lechner1992}
R.~E. Lechner,
\newblock {\em Optimization of the chopper system for the cold-neutron
  time-of-flight spectrometer NEAT at the HMI, Berlin},
\newblock Physica B {\bf 181}, 973--977 (1992).

\bibitem{Lechner1996}
R.~E. Lechner, R.~Melzer, and J.~Fitter,
\newblock {\em First QINS results from the TOF-spectrometer NEAT},
\newblock Physica B {\bf 226}, 86--91 (1996).

\bibitem{Schober2000}
H.~Schober, A.~J. Dianoux, J.~C. Cook, and F.~Mezei,
\newblock {\em Upgrade of the IN5 cold neutron time-of-flight spectrometer},
\newblock Physica B {\bf 276-278}, 164--165 (2000).

\bibitem{Ollivier2002}
J.~Ollivier, H.~Casalta, H.~Schober, J.~C. Cook, P.~Malbert, M.~Locatelli,
  C.~Gomez, S.~Jenkins, and I.~J. Sutton,
\newblock {\em New perspectives on the IN5 time of flight spectrometer},
\newblock J Appl Phys A {\bf 74}, S305--S307 (2002).

\bibitem{Ollivier2004}
J.~Ollivier, M.~Plazanet, H.~Schober, and J.~C. Cook,
\newblock {\em First results from the upgraded IN5 disk chopper cold
  time-of-flight spectrometer},
\newblock Physica B {\bf 350}, 173--177 (2004).

\bibitem{Copley2003a}
J.~R.~D. Copley and J.~C. Cook,
\newblock {\em The Disk Chepper Spectrometer at NIST: a new instrument for
  quasielastic neutron scattering studies},
\newblock Chem Phys {\bf 292}, 477--485 (2003).

\bibitem{Copley2003b}
J.~R.~D. Copley,
\newblock {\em An acceptance diagram analysis of the contaminant pulse removal
  problem with direct geometry neutron chopper spectrometers},
\newblock Nucl Instr and Meth Phys Res A {\bf 510}, 318--324 (2003).

\end{thebibliography}
  \bibliographystyle{citations_style}

\end{document}